\documentclass[fleqn,twoside]{article}
\usepackage{espcrc2,epsfig}

\usepackage{graphicx}
\usepackage[figuresright]{rotating}

\newcommand{\nn}{\nonumber}
\newcommand{\beq} {\begin{equation}}
\newcommand{\eeq} {\end{equation}}
\newcommand{\beqa} {\begin{eqnarray}}
\newcommand{\eeqa} {\end{eqnarray}}

\newcommand{\as}{\alpha_s}

\newcommand{\ieps}{i\varepsilon}
\newcommand{\order}[1]{${\cal O}\left(#1 \right)$}

\newcommand{\eq}[1]{(\ref{#1})}
\newcommand{\ket}[1]{\vert{#1}\rangle}
\newcommand{\bra}[1]{\langle{#1}\vert}
\newcommand{\ave}[1]{\left\langle{#1}\right\rangle}

\newcommand{\pvec}{\vec p}

\newcommand{\ol}{\overline}     

\newcommand{\Slash}[1]{ \parbox[b]{0.6em}{$#1$} \hspace{-0.55em}
                            \parbox[b]{0.55em}{ \raisebox{-0.2ex}{$/$}}}

\title{\vskip -30pt %
{\hbox to\hsize{\normalsize\hfil\rm  26 September, 2002}} \vskip -3pt
{\hbox to\hsize{\normalsize\hfil\rm HIP-2002-44/TH}}
Perturbative QCD with Quark and Gluon Condensates\thanks{Talk at ICHEP 2002, Amsterdam (July 2002).}}

\author{Paul Hoyer\address{Department of Physical Sciences and Helsinki Institute of Physics\\ 
        P.O. Box 64, FIN-00014 Helsinki University, Finland}%
        \thanks{Address until 1 August 2002: Nordita, Blegdamsvej 17, DK-2100 Copenhagen, Denmark. Research supported in part by the
European Commission under contract HPRN-CT-2000-00130.}}
       
\begin{document}

\begin{abstract}
The quark $\ave{\bar\psi \psi}$ and gluon $\ave{F_{\mu\nu}F^{\mu\nu}}$ vacuum expectation values are non-vanishing at low orders in $\as$ when the perturbative ground state includes quark and gluon pairs. This offers possibilities of studying the effects of a non-trivial vacuum using formally exact perturbative methods. The QCD Feynman rules are modified only for the free, on-shell quark and gluon propagators at zero four-momentum. Gauge and lorentz invariance is maintained, while chiral symmetry is spontaneously broken by the ground state.
\vspace{1pc}
\end{abstract}

\maketitle 

\section{THE GROUND STATE}

The perturbative expansion is the basis of most comparisons of gauge field theories with data, and is used to establish general properties like analyticity and factorization of scattering amplitudes. The expansion is determined by the lagrangian ${\cal L}$ together with
\begin{itemize}

\item[(a)] The renormalization scheme $(Q^2 \to \infty)$

\item[(b)] The perturbative ground state $(Q^2 \to 0)$

\end{itemize}
While the dependence on the short-distance regularization (a) has been thoroughly studied, I find it surprising that little attention is paid to the starting point (b) of the expansion. The essential difference between QED and QCD is believed to lie in the nature of the QCD vacuum. The poor approximation provided by the empty perturbative vacuum of the quark and gluon condensates is a potential reason that perturbative QCD (PQCD) fails to describe long-distance phenomena.

It may be argued that PQCD is in any case inadequate at long distance due to the large size of the strong coupling $\as$. While this possibility cannot be excluded, it is not a sufficient reason for neglecting to consider alternative choices of the perturbative vacuum. PQCD with a non-empty ground state could serve as a model to improve our understanding of vacuum effects (including confinement). Furthermore, analyses of data \cite{MovillaFernandez:2001ed} indicate that the strong coupling freezes at a value $\as(Q^2 = 0) \simeq 0.5$. At a more qualitative level, the hadron spectrum shares many features with the hydrogen atom. Thus quark degrees of freedom and PQCD appear to be relevant in the confinement region \cite{yd} -- but attention needs to be paid to the choice of ground state.

The perturbative expansion is based on using free $\ket{in}$ and $\bra{out}$ states at asymptotic times, $t=\pm \infty$. Formally, any choice of asymptotic state that has an overlap with the true ground state is allowed, since the state will relax to the ground state during its evolution to finite $t$. At zeroth order in $\as$ the free asymptotic state is an eigenstate of the hamiltonian and thus constitutes the perturbative ground state at all times.

In path integral derivations of the perturbative expansion the choice of an empty perturbative vacuum is usually made implicitly. The dependence on the wave function of the asymptotic state can be made explicit by first considering a finite time interval, $-T \leq t \leq T$, and then taking the $T \to \infty$ limit \cite{Hoyer:2002ru}. Thus the generating functional of a free scalar field (in 0+1 dimensions, to keep the notation simple)
$$
Z_0[J] = \int {\cal D}[\phi] \exp\left\{i\int dt \left[{\cal L}_0(\phi) + J(t)\phi(t)\right]\right\}
$$ 
becomes
\beqa \label{zfree}
Z_0[J] = \int {\cal D}[\phi(t)] \exp\left\{\frac{-i}{2}\int_{-T}^T dt\, \ \ \ \ \ \ 
\right. && \nn \\
\phi(t)\bigg[\frac{\partial^2}{\partial t^2}+ m^2 \bigg] \phi(t) + i\int_{-T}^T dt\, J(t)\phi(t) && \nn \\
 -\left. \frac{C_s}{2}\Big[\phi^2(-T)+\phi^2(T)\Big]\right\} &&
\eeqa
where the path integral is over $\phi(-T\leq t \leq T)$. The wave functions at $t=\pm T$ should not include interactions and thus are gaussian. The ground (empty) state of this harmonic oscillator system has $C_s = m$. Other choices of $C_s$ can be seen as superpositions of states with particle pairs through an expansion in powers of $(C_s-m)\phi^2$.

In the $T\to \infty$ limit the free generating functional \eq{zfree} becomes
\beqa \label{bosres}
Z_0[J] =\exp\left\{-\frac{1}{2}\int_{-\infty}^\infty \frac{dE}{2\pi} J(-E) \left[\frac{i}{E^2-m^2+\ieps} 
\right.\right. && \nn \\ \left.\left. 
+ \left(\frac{C_s}{m}-1\right) \pi\delta(E^2-m^2)\right]J(E)\right\}
\ \ (2) && \nn
\eeqa \addtocounter{equation}{1}
Thus only the on-shell part of the free propagator is modified when $C_s \neq m$. This is natural since on-shell propagation is required to feel the boundary conditions at $t=\pm \infty$. It is also important to note that the generating functional of the interacting theory,
\beq \label{zint}
Z[J]= \exp\left[iS_{int}\left(\frac{\delta}{i\delta J}\right)\right] Z_0[J]
\eeq
defines the feynman rules in terms of standard interaction vertices and the free propagator of Eq. \eq{bosres}. Hence the perturbative expansion is modified {\em only} by the additional term $\propto C_s-m$ in the free, on-shell propagator.

The above method can be trivially extended to 3+1 dimensions. The freedom in choosing asymptotic states allows the addition of pairs with non-vanishing relative three-momentum \cite{Hoyer:2000ca}. Lorentz symmetry is restored due to the relaxation of the asymptotic state to the (lorentz invariant) ground state. In this case the perturbative expansion will not, however, be lorentz invariant order by order. Adding pairs only for massless fields with $\pvec = 0$ retains lorentz invariance order by order in $\as$ and is thus an attractive option \cite{Cabo:1995za}. 

Similar considerations show \cite{Hoyer:2002ru,Cabo:1995za} that the PQCD expansion is modified only through the on-shell quark and gluon propagators,
\beqa
S_q^{AB}(p) = \delta^{AB}\left[\frac{i\Slash{p}}{p^2+\ieps}+C_q (2\pi)^4
\delta^4(p)\right] &&
\label{qpropmod}\\
D_g^{ab,\mu\nu}(p) = -g^{\mu\nu} \delta^{ab}\left[\frac{i}{p^2+\ieps}+C_g
(2\pi)^4 \delta^4(p) \right] && \label{gpropmod}
\eeqa
The constants $C_q$ and $C_g$ determine the density of $q\bar q$ and gluon pairs in the perturbative ground state and are the only parameters allowed by this generalization of PQCD. The corresponding modification of the ghost propagator is irrelevant since the ghost-gluon vertex is proportional to the ghost momentum.

\vspace{-1cm}
\begin{figure}[hbt]
\begin{center}
\leavevmode
{\epsfxsize=7truecm \epsfbox{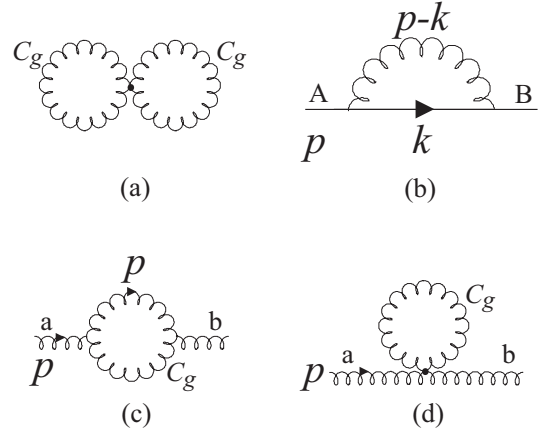}}
\end{center} \vspace{-1cm}
\caption[*]{(a) The only diagram which contributes to $\ave{F_{\mu\nu}F^{\mu\nu}}$ at \order{g^2}. The gluon condensate term $\propto C_g$ in \eq{gpropmod} is taken in both loops. (b) Quark propagator correction. (c,d) Gluon propagator corrections.} \vspace{-.5cm}
\label{fig1con}
\end{figure}

\section{PHYSICAL EFFECTS}

The modifications \eq{qpropmod}, \eq{gpropmod} of the quark and gluon propagators change the analytic structure of scattering amplitudes. This reflects scattering on the on-shell partons in the perturbative ground state. Such scattering also affects quarks and gluons propagating in the condensates of the true QCD vacuum. PQCD with quark and gluon pairs in the perturbative vacuum may thus serve as a model of QCD vacuum effects in feynman diagrams. It is of particular interest to find the eigenstates of propagation in the perturbative vacuum, which can serve as external states in the S-matrix.

The quark propagator \eq{qpropmod} gives a finite quark condensate already at zeroth order in perturbation theory,
\beq \label{qcond}
\ave{\ol{\psi}(x)\psi(x)} = -4NC_q
\eeq
where $N$ is the number of colors. Thus chiral symmetry is spontaneously broken by the asymptotic state.

In the case of the $\ave{F_{\mu\nu}F^{\mu\nu}}$ expectation value the \order{g^0} contribution vanishes since the free part of $F_{\mu\nu}=\partial_\mu A_\nu -\partial_\nu A_\mu -g f_{abc}A_\mu^b A_\nu^c$ is proportional to the gluon momentum. Furthermore, at \order{g^2} the gluon loop integrals vanish (when dimensionally regularized) due to the absence of a mass or momentum scale. This leaves only the two-loop diagram involving a 4-gluon coupling (Fig. 1a), with the $C_g$ part of the propagator taken in both loops. The result is \cite{Cabo:1995za}
\beq \label{gcond}
\ave{F_{\mu\nu}(x)F^{\mu\nu}(x)} = 12g^2 N(N^2-1)C_g^2
\eeq
The gluon condensate thus arises specifically from the non-abelian 4-gluon interaction.

The quark propagator correction shown in Fig. 1b gets a contribution from both the quark condensate term $C_q$ in \eq{qpropmod} ($k=0$) and from the gluon propagator modification $C_g$ of \eq{gpropmod} ($k=p$). Only the former breaks chiral symmetry and thus generates a proper `constituent' quark mass \cite{Hoyer:2002ru},
\beq \label{constmass}
M_q = \Big(4g^2C_F|C_q|\Big)^{1/3}
\eeq
which then gets a correction from the gluon condensate term $C_g$.

The sum of the gluon propagator corrections shown in Figs. 1c,d has the transverse structure required by gauge invariance. At order $g^2$ the gluon mass
\beq \label{gluemass}
M_g^2 = -2g^2 N C_g
\eeq
is tachyonic \cite{Hoyer:2000ca,Cabo:1995za} if $C_g>0$.

\begin{figure}[hbt]
\begin{center}
\leavevmode
{\epsfxsize=7truecm \epsfbox{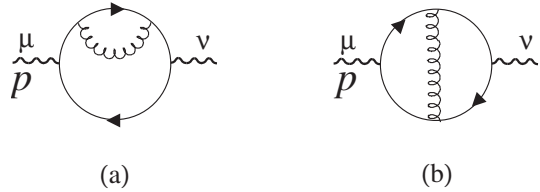}}
\end{center} \vspace{-1cm}
\caption[*]{Corrections to the color singlet (axial) vector current. A third diagram where the gluon couples to the antiquark is not shown.} \vspace{-.5cm}
\label{fig1con}
\end{figure}

The present framework allows a perturbative demonstration of how the pion (goldstone) mass vanishes in spite of the constituent quark mass generated by corrections like Fig. 1b. The vanishing of the pion mass is a consequence of the general relation
\beq \label{axpion}
\bra{0}J_5^\mu(x)\ket{\pi(p)} = -ip^\mu f_\pi \exp(-ip\cdot x)
\eeq
and the conservation of the axial vector current, $\partial_\mu J_5^\mu(x) =0$ when $m_q=0$. The \order{g^2} quark mass correction to the divergence of the axial vector propagator shown in Fig. 2a is exactly cancelled by the quark interaction term of Fig. 2b \cite{Hoyer:2002ru}.

\vspace{.5cm}
{\bf Acknowledgments.} I am grateful for helpful discussions with St\'ephane Peign\'e.

\end{document}